# Electron beam transfer line design for plasma driven Free Electron Lasers

M. Rossetti Conti[a,b,d], A. Bacci[a], A. Giribono[c], V. Petrillo[b], A.R. Rossi[a], L. Serafini[a], C. Vaccarezza[c]

[a] INFN - MI, via G. Celoria 16, 20133 Milan, Italy
[b] University of Milan, Physics Department, via G. Celoria 16, 20133 Milan, Italy
[c] INFN - LNF, v.le E. Fermi, 00044 Frascati, Italy
[d] Corresponding author: marcello.rossetti@mi.infn.it

**ABSTRACT**

Plasma driven particle accelerators represent the future of compact accelerating machines and Free Electron Lasers are going to benefit from these new technologies. One of the main issue of this new approach to FEL machines is the design of the transfer line needed to match of the electron-beam with the magnetic undulators. Despite the reduction of the chromaticity of plasma beams is one of the main goals, the target of this line is to be effective even in cases of beams with a considerable value of chromaticity. The method here explained is based on the code GIOTTO [1] that works using a homemade genetic algorithm and that is capable of finding optimal matching line layouts directly using a full 3D tracking code.

**INTRODUCTION**

The Free Electron Lasers (FEL), tunable lasers classified as $4^{th}$ generation light sources, have spread widely in the most advanced research centers since their first theorization by John Madey in 1971 at Stanford University [2].This success is for sure due to their tenability and to the possibility to use these machines as sources of coherent light in the UV frequencies and X-rays range, where conventional lasers are not available. These X-ray coherent radiation is very important for bio-medical analysis, materials scanning and more others frontier applications [3].
Nowadays the performances of plasma acceleration experiments are pushing forward, thanks to the high interest of the whole scientific community, and this makes it desirable an application to get more compact FEL machines. A compact layout can be achieved using a very short plasma stage as main accelerator and building a compact transfer line able to control and match the beam coming from the plasma acceleration stage in the undulator section.

We studied the possibility of application of a Genetic Algorithm to the matching problem of a machine like the Eupraxia@SPARCLAB Free Electron Laser. We used a Genetic Code for beam-dynamics optimizations named GIOTTO and a tracking algorithm called ASTRA [4] to perform a Wide Range Search (following referred as WRS) for the correct parameters of the Transfer Line (TL).

In this work are presented 4 different TLs lattices found using GIOTTO. These lines are different in terms of lengths, number of quads and type of quads (permanent or electromagnetic). We did not enter in merit of the different advantages or disadvantages between the different lines (as jitters or misalignments), because the aim of the paper is to present a new methodology to cope with this kind of problems.

We obtained, by means of simulations, a laser driven plasma beam pre-accelerated by a conventional accelerator that was able to produce an ad hoc electron bunch (witness) for the external injection in a laser driven plasma wave (Laser Wake Field Accelerator, LWFA). The conventional accelerator was tuned in a peculiar way: we simulate a velocity bunching scheme in the first accelerating cavity and then we tune the RF focusing in the booster. With this strategy we transport the bunch up to the transfer line entrance without breaking its own cylindrical symmetry, which means to preserve the minimum bunch emittance.

In this proceeding I will introduce the concept of Chromatic Length and why it is so important to take it into account in plasma acceleration experiments. I will then briefly show the concept of matching the beam in the undulator modules and how it fixes the beam characteristics that we have to obtain with our transfer line. I will show then why we chose a genetic algorithm to design and test the transfer line stressing out some improvements of GIOTTO we made, then I will present four alternative transfer lines obtained with this methodology that are able to match the given bunch. A small focus on the most interesting line will be shown in conclusion.

**THE CHROMATIC LENGTH**

Designing a transport line for plasma accelerated bunches, able to preserve beam properties, requires particular care [5]. It has been shown that a relatively high amount of energy spread results in a consistent normalized emittance dilution even in drifts [6], whence the widespread conviction for the need of putting beam capturing optical elements as close as possible, compatibly with beam line constraints, to the plasma accelerator exit. However, an exact quantification of "how close" has never been given, so that beam line designers cannot decide whether optics are close enough or this requirement can eventually be relaxed in their specific situation. In the following, we would like to provide an answer to this pressing question.

Following [6] we write the normalized emittance as:

$$\varepsilon_n^2 = \langle \gamma \rangle^2 (\sigma_E^2 \sigma_x^2 \sigma_{x'}^2 + \varepsilon_0^2) \tag{1}$$

Where $\langle \gamma \rangle$ is the average Lorentz factor, $\sigma_E$ the relative energy spread, $\sigma_x$ and $\sigma_{x'}$ respectively the values of bunch transverse size and divergence and $\varepsilon_0$ its initial geometric emittance $\varepsilon_0^2 = (\sigma_x^{(0)})^2(\sigma_{x'}^{(0)})^2 - (\sigma_{xx'}^{(0)})^2$. The beam transverse size evolution, assuming an emittance dominated beam, is $\sigma_x^2(s) = (\sigma_{x'}^{(0)})^2 s^2 + \sigma_{xx'}^{(0)} s + (\sigma_{x'}^{(0)})^2$, where $s$ is the drift length. Inserting this result into (1) and requiring the first term in parenthesis to be equal to the second (i.e. requiring a doubling of the initial squared normalize emittance), yields a drift length of

$$s = -\frac{\sigma_{xx'}^{(0)}}{(\sigma_{x'}^{(0)})^2} \pm \frac{\varepsilon_0 \sqrt{1-\sigma_E^2}}{\sigma_E (\sigma_{x'}^{(0)})^2}$$

Considering the positive solution, assuming the bunch initial condition is a waist (as is typical for a beam coming out from a plasma channel, disregarding ramps) and $\sigma_E^2 \ll 1$, allows to define the chromatic length as

$$L_C \cong \frac{\varepsilon_0}{\sigma_E (\sigma_{x'}^{(0)})^2} = \frac{\sigma_x^{(0)}}{\sigma_E \sigma_{x'}^{(0)}} = \frac{(\sigma_x^{(0)})^2 \cdot \gamma}{\sigma_E \cdot \varepsilon_0} \tag{2}$$

By definition, $L_C$ has a meaning equivalent to the Rayleigh length for a laser beam or the diffraction length for an emittance dominated particle beam, namely signals when the relevant quantity increases by a factor $\sqrt{2}$ with respect to its initial value or, equivalently, a range within which it can be considered approximately constant. It also explicitly shows how normalized emittance dilution is driven both by high energy spread and a small ratio between beam size and divergence.

For a typical linac accelerated bunch with several hundreds of MeV energy and μm level normalized emittance, $\sigma_x^{(0)} \sim 10\ \mu m$, $\sigma_{x'}^{(0)} \sim 100\ \mu rad$ and $\sigma_E \lesssim 10^{-3}$ gives $L_C \gtrsim 10^2\ m$ which is usually longer than the accelerator itself whereas, for a typical plasma accelerated beam of the same energy and emittance, $\sigma_x^{(0)} \sim 1\ \mu m$, $\sigma_{x'}^{(0)} \sim 1\ mrad$ and $\sigma_E \gtrsim 10^{-2}$, the chromatic length is reduced to $L_C \lesssim 10^{-1}\ m$, pointing out the need from putting any optical element as close as possible (i.e. within $L_C$) to the plasma channel end.

## THE MATCHING IN UNDULATOR

The trajectory of an electron bunch in a long undulator is unstable since this element develop quadrupolar components of its magnetic field. In fact, this element is characterized by a strongly vertical focusing effect and a weak horizontal defocusing effect on the electron beam. For this reason, the undulators are divided in modular portions separated by magnetic quadrupoles that compensate the undulator quadrupole effect.

The matching in the undulator is performed imposing the periodicity of the Twiss functions (α and β) all over the periodic module of the lattice. Once the undulator is chosen, the existence of the solution of this problem depends on beam parameters, such as average beam energy (and normalized transverse emittance), and on the emitted radiation wavelength [7].

It is also very important to crosscheck the quadrupole strength needed to compensate the undulator unwanted effects and the spot size of the beam in the quadrupole bore, the values of these parameters must be in agreement with the state of the art of the technology of the quadrupoles. Once this condition is granted, the exact values of the Twiss functions of the bunch at the entrance of the first undulator module are known and the design of the transfer line, i.e. the set of quadrupoles upstream the undulator that are responsible of matching these initial parameters (used below as target parameters of our genetic code), can start.

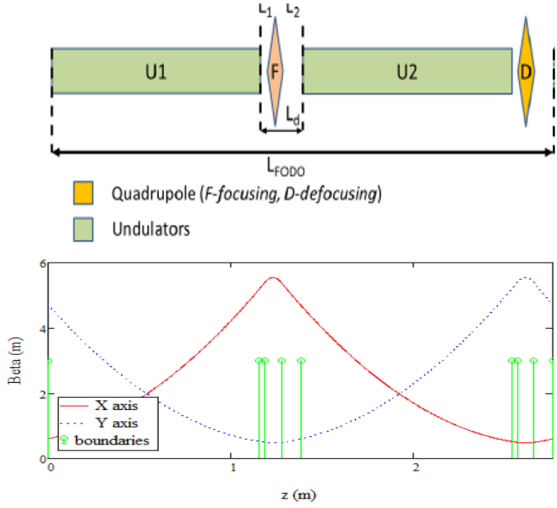

**Figure 1**. Up. The periodic block of the undulator of Eupraxia@SPARCLAB, composed by 2 undulators and 2 quadrupoles. The quadrupoles are in FODO configuration. Down. The values of the transverse β functions in the periodic block. This setting of the line grant the periodicity of the solution all over the undulator if the beam is matched.

**DESIGNING A TRANSFER LINE**

The beamline portion dedicated to match the beam at the entrance of the first undulator is called Transfer Line (TL) and the magnetic elements used in this portion are quadrupoles. In fact, as said before, at the entrance of the first undulator module it is needed to obtain values of the Twiss functions as close as possible to the ones that grants the periodicity of these functions in the modules.

The TL lattice is often studied analytically. The quadrupoles and the drifts are described by their associated transfer matrix, like lenses in optics. The product between the 6D vector associated with the beam

$X = (x, p_x, y, p_y, \Delta z, \delta)^T$ and these matrices give the transferred beam vector after the propagation through the element. From the transfer line equivalent matrix, obtained multiplying all the matrices associated with the optic elements, one can obtain the transfer matrix for the Twiss parameters [8] and figure out how to design the TL.

Considering the layout of the machine, the starting point of the TL will coincide with the output of the plasma channel, here the bunch suddenly passes from a region characterized by an extremely strong electric field to a drift space. The beam inherits from the plasma acceleration stage a big energy spread and transverse momenta. In these special conditions most of the approximations done to obtain the transfer matrices are lost and it's necessary to consider the quadrupole transfer matrices approximated up to the second order [9]. The analytical approach become more complex with solutions which have to be carefully checked by simulation. Considering these problematics, we have chosen a different approach which is based directly on the particle tracking able to give a full phenomenon description.

As said before, the propagation inside the plasma channel is responsible of the high values of the beam energy spread and transverse momenta, the combination of these features with the small transverse dimensions of the bunch at the beginning of the TL causes big issues due to the high chromaticity of the beam.

Let us look again at the Chromatic Length defined before for a transversally symmetric beam ($\sigma_x^{(0)} \cong \sigma_y^{(0)}$), equation (2).

For the bunch that we obtained, described in the section "Beam Preparation" and shown in Fig. 2, with transverse rms dimensions $\sigma_x \cong \sigma_y \cong 1.8 \cdot 10^{-5}$ m, energy spread $\sigma_E \cong 2.7 \cdot 10^{-2}$ and divergence $\sigma_{x'} \cong \sigma_{y'} \cong 0.14$ mrad, the Chromatic Length results $L_C \cong 4.8$ m. The rms transverse dimension value is strongly affected by a large halo in the tail due to the interaction with the plasma. We performed a transverse cut on the full beam distribution, with radius equal to $1 \cdot 10^{-5}$ m, to avoid a misleading growth of the rms beam transverse dimensions when transporting the full beam through the line.

Focusing again on the methodology to solve the design of TLs, in the next chapter we are going to explain the procedure we chose to adopt, able to give the TL lattice solution directly from tracking code simulation results. The main actor in our procedure is a genetic algorithm able to generate TLs lattices, drive a tracking code to analyze the goodness of the proposed solutions and then to create better lattices than previous ones.

**A GENETIC ALGORITHM TO DESIGN A TRANSFER LINE**

The code used for the TL parameters search is GIOTTO [1], an up-to-date Genetic Algorithm (GA), fully developed in our group since 2007, written in Fortran 90 and able to perform statistical analysis [10] and genetic optimizations of a beamline [11][12]. This is possible because the software is able to drive and to communicate with a tracking code, ASTRA [4], through its input and output files.

GA are a class of stochastic optimizers that deal very well with problems in which the parameters values are correlated in a strongly nonlinear way. The particles beam dynamics in a beamline can be a problem with these characteristics. For example, at low beam energy the space charge turn-on many nonlinear correlations which usually make very complex the machine tuning. Also at high energy very strong longitudinal compression factor, or transversal focusing, as nowadays requested in many application, can turn-on again the space charge non-linear effects. Further the high chromaticity typical of plasma accelerated beams makes it very complex to design matching lines by the non-conservation of the normalized emittance. As shown above by the chromatic length.

Our approach to the design of a TL is to start from scratch, i.e. defining a chosen set of switched off quadrupoles (with focusing strengths equal to zero) with arbitrary positions, and change the strengths and the positions gradually, evaluating the fitness value for each solution. Let us compare this problem with a working point optimization problem, in which the main behavior of the line is known and the values of the parameters that identify the solution are in the neighborhood of the values of the starting point of the optimization. In our case the GA need to perform in a bigger area a WRS for the correct parameters. In fact, GA are able to deal very well with this different kind of optimization problems thanks to their ability to move wisely in the space of the solutions. A classical didactic example of application of GA is on the Travelling Salesmen Problem (TSP) that requires to start from scratch and move in a wide solution space [13].

For these reasons we needed to develop furthermore GIOTTO in order to deal better with low values of portions of the fitness function (when the solution is close to some target values but far from others) and to change the way we define the fitness score of a solution. In particular, we worked on the fitness function shape, looking for a distribution which was not too much severe with values far to the target (which means to work on a large domain). The fitness function we used (shown below) is a sum of Lorentzian functions, six in this case, each one centered on the target value of the parameters that we want to optimize at the end of the TL, i.e. the transverse Twiss functions needed for the beam matching to the undulator and the normalized transverse emittances that must be kept as low as possible:

$$I = \sum_{i=1}^{6} A_i \frac{B_i^2}{B_i^2 + (x_{Ti} - x_{Fi})^2}$$

Where $I$ stays for idoneity score (or fitness score), $x_{Ti}$ are the target values (shown in Table 1) of the six parameters we chose to optimize, $x_{Fi}$ are the final values of these parameters that are obtained at the end of the TL, $A_i$ and $B_i$ are coefficients of the Lorentzian functions that determinate respectively height and width of the curves.

GIOTTO can work with every analytical fitness functions; defined directly by the user. In previous optimizations, here not considered, we often used Gaussian functions, which are not indicate for WRS problems, as the one here discussed. The point is that the slope of the Gaussian tails goes down very fast, moving out of the optimization a big portion of the population, which should contribute to the evolution. Lorentzian functions, compared to the Gaussian, show a much higher slope values also for far tails. This guarantee the ability to find solutions even in WRS problems. The height of the Lorentzian can be adjusted to change the priority of the target (or object) optimization, the width is used to change the radius of the neighborhood of the target value in which the object value is trapped. This gives to Giotto the possibility to works in a Multiobjective Genetic Algorithms way MOGA [14]

The final parameters that we needed to control in our optimization are the four values of the transverse Twiss functions $\alpha_x, \alpha_y, \beta_x$ and $\beta_y$ that must match the values that grant the periodicity condition in the undulator and the values of the transverse normalized emittances of the beam $\varepsilon_{n,x}$ and $\varepsilon_{n,y}$. All these parameters are evaluated at the end of the TL. It is important to control the emittance and keep it constant (or quasi constant) avoiding solutions where the Twiss parameters are correct, but with a different emittance value. This emittance variation, for highly chromatic beams, is trigged by strong defocusing of the particles (a chromatic aberration).

| | |
|---|---|
| $x_{T1} = \alpha_{xT}$ | 1.48 |
| $x_{T2} = \beta_{xT}$ | 5.04 m |
| $x_{T3} = \alpha_{yT}$ | -0.65 |
| $x_{T4} = \beta_{yT}$ | 2.11 m |
| $x_{T5} = \varepsilon_{n,xT}$ | 0.42 mm mrad |
| $x_{T6} = \varepsilon_{n,yT}$ | 0.43 mm mrad |

**Table 1**: Target values of the parameters that are optimized in the algorithm, a Lorentzian function is assigned to each parameter.

**PROCEDURE AND RESULTS**

**Beam preparation.** A 40 pC high brightness electron beam have been obtained by simulation using the tracking code ASTRA. The Linac layout simulated was optimized in term of compression, emittance preservation and RF focusing, with the aim to obtain beam parameters ad hoc for the injection into a plasma wave. The beam is than furthermore focused with a quadrupole triplet and then accelerated up to 1 GeV of energy with a PWFA simulation (with Qfluid [15]).

A slice analysis of the bunch at the extraction point of the plasma is performed in order to select the portion of the beam with the best quality, in terms of brightness (highlighted in Fig. 2), and match this portion with the TL we are going to design.

A cut of the beam is performed to select the particles that we want to exploit to produce the radiation (Fig. 2, Table 1) The slice parameters are then used to check the existence of a periodic solution of the Twiss functions associated with this slice inside the undulator. This operation has been performed with GENESIS [16] and the resulting initial values of the Twiss functions are shown in Table 2.

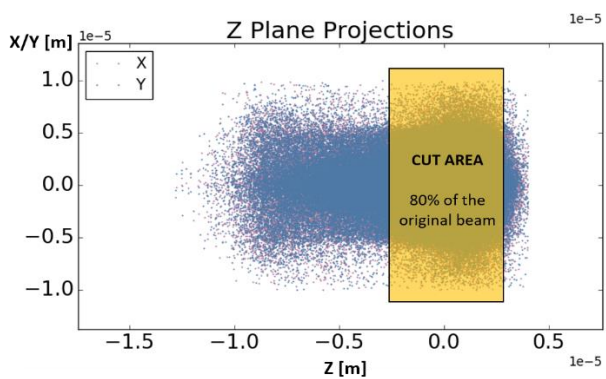

**Figure 2**: The portion of the beam that have been matched, we chose to select the portion with the highest peak current and lowest emittance; in this figure a large beam halo situated at the height of the tail is not shown to show better the cut area.

| | |
|---|---|
| $\sigma_x$ | 2.4 μm |
| $\sigma_y$ | 2.4 μm |
| $\varepsilon_{n,x}$ | 0.42 mm mrad |
| $\varepsilon_{n,y}$ | 0.43 mm mrad |
| Energy | 1 GeV ($\gamma \cong 2000$) |
| $\sigma_E$ | $3.29 \cdot 10^{-3}$ |
| Total Charge | 32 pC |

**Table 2**: Cut beam parameters used for the matching to the undulator.

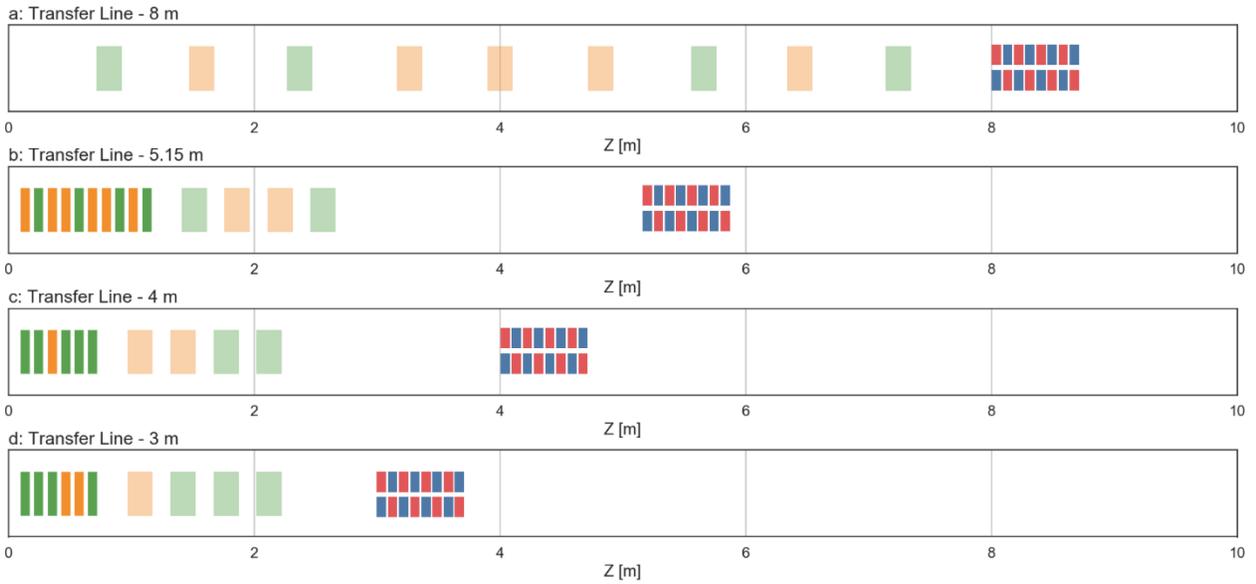

**Figure 3**: schematic lattice of 4 alternative TLs obtained for different line lengths (from top to bottom: 8 m, 5.15 m, 4 m, 3 m). The orange and the green elements are quadrupoles, with focusing and defocusing effect on the horizontal plane. The transparent style means that they are EMQ, the colourful style mean that they are PMQ. In red and blue is shown (not in scale) the position of the first undulator module.

**TL optimization and results.** First, we checked how to reduce the tracking time of the beam in the TL without introducing artifacts in the simulation results. We found that using $10^5$ macroparticles to simulate the bunch and disabling the space charge effects is the best compromise. Once the optimization gives us a good solution (i. e. a set of positions and focusing strengths), a tracking with more particles ($2 \cdot 10^6$) and space charge effects taken into account is done to verify the goodness of the line.

With this procedure, we obtained four different possible TLs, they differ in length and quadrupole magnets technology (Fig. 3). All the lines make use of Electro-Magnetic Quadrupoles (EMQ), the shorter lines (B, C and D) take advantage of

Permanent Magnet Quadrupoles (PMQ). PMQ can generate much stronger focusing gradients, up to 700 T/m vs 70 T/m of maximum gradient for EMQ.

In Table 3 there is a summary of the 4 TLs obtained with the GA, they are all able to match the bunch, the final Fitness values differs from its theoretical maxi-mum by less than 1‰, and a summary of the focusing strengths is given in Table 4.

|   | Length | # EMQ | EMQ length | EMQ bore | # PMQ | PMQ length | PMQ bore |
|---|---|---|---|---|---|---|---|
| A | 8 m | 9 | 20 cm | 1 cm | | | |
| B | 5.15 m | 4 | 20 cm | 1.5 cm | 10 | 7 cm | 1cm |
| C | 4 m | 4 | 20 cm | 1.5 cm | 6 | 7 cm | 1cm |
| D | 3 m | 4 | 20 cm | 1.5 cm | 6 | 7 cm | 1cm |

**Table 3**: Magnetic elements in the 4 different TLs shown in Fig. 3

|   | Q1 | Q2 | Q3 | Q4 | Q5 | Q6 | Q7 | Q8 | Q9 | Q10 | Q11 | Q12 | Q13 | Q14 |
|---|---|---|---|---|---|---|---|---|---|---|---|---|---|---|
| A | -14 | 10 | -11 | 8.7 | 8.7 | 4.3 | -8.2 | 6.1 | -8.3 | - | - | - | - | - |
| B | **49** | **-134** | **35** | **155** | **-158** | **170** | **85** | **-47** | **78** | **-53** | -5.9 | 8.3 | 7.1 | -7. |
| C | -12 | **-120** | **100** | **-6** | **-19** | **-2** | 6.7 | 0.5 | -5.2 | -4.1 | - | - | - | - |
| D | -68 | -21 | -61 | **77** | **18** | **-31** | 2.1 | -3.0 | -3.8 | -4.0 | - | - | - | - |

**Table 4**: Focusing strengths of the four solutions; in bold for PMQ. The values are rounded for sake of compactness; the aim is to give an idea of their magnitude.

In these optimizations, we did not define constraints on quads positions and the number of quads was fixed. A quads overlapping means the necessity to revise the lattice, which never happened in these cases. The final positions found are in agreement with a feasible installation and can be deduced on the Fig. 3 (in scale). We don't consider here necessary to show all these data, consider this work a proof of principle of the goodness of the method.

Looking at Table 4 it's easy to notice that, in some cases, consecutive elements have the same sign of the field, it is due to the number of elements initially chosen, the mechanics of the GA will generate a solution that use all the elements. This condition may suggest the possibility to substitute them with an equivalent longer element, this can be done to obtain a simpler line but with less knobs for possible future change of working points. At this point the choice of the strategy to be adopted is linked to one's own needs. In this case, for example, we were looking for a line that could be furthermore optimized in focusing strengths to match to the undulator beams with different energies (see below). We preferred a more complex, but also more flexible, line with a higher number of elements precisely for this reason.

The line C demonstrated to be a good compromise between compactness and reduced growth of the Twiss functions in the PM portion of the line.

**TL flexibility.** The C line option have been tested to be able to operate also with beams of significantly different energy. We assumed to use some tunable PMQ that are being developed in these years [17] and repeated the optimization for the same beam but changing its average energy to 500 MeV in the first case and to 2 GeV in the second case and changing the sets of the quads strengths. The gradients was moved inside a range compatible with the state of the art of PMQ/EMQ capabilities. The TL demonstrated to deal very well with this strong changes in the beam parameters. In fact, these quadrupoles positions showed to be able to keep the envelopes dimensions under control preventing beam losses. A visual comparison on the envelopes in the line C for 500 MeV, 1 GeV and 2 GeV is provided in Fig. 4.

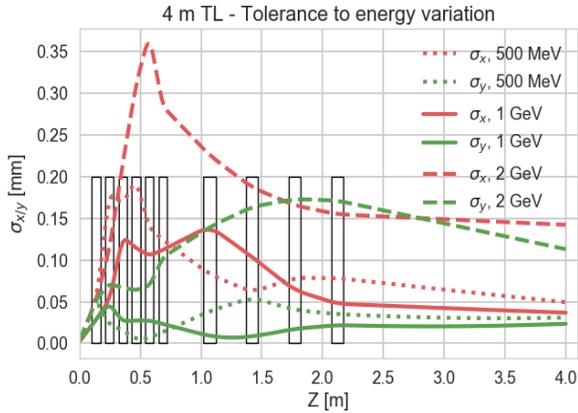

**Figure 4**. Comparison between the beams envelopes for matched bunches of significantly different energies. The position and the dimension of the elements of the line is the same in the three cases (rectangles), only the quads strengths are changed.

|  | Q1 | Q2 | Q3 | Q4 | Q5 | Q6 | Q7 | Q8 | Q9 | Q10 |
|---|---|---|---|---|---|---|---|---|---|---|
| 0.5 GeV | **-205** | **116** | **-33** | **61** | **-40** | **-1.6** | -2.8 | -8.5 | 3.3 | 0.9 |
| 1 GeV | **-12** | **-120** | **100** | **-6** | **-19** | **-2** | 6.7 | 0.5 | -5.2 | -4.1 |
| 2 GeV | **-54** | **-46** | **51** | **15** | **54** | **-31** | -0.9 | -1.3 | -0.9 | -0.7 |

**Table 5**: Focusing strengths of the line C for different energies; in bold for PMQ. The values are rounded for sake of compactness; the aim is to give an idea of their magnitude.

## CONCLUSION

In this paper, it has been shown how to apply the code GIOTTO [1] (a GA) on the TL design problem to perform the matching to the undulator, in particularly when electron beams suffer of strong chromatic effect that quickly degrade the emittance, which is the case of beams coming from plasma accelerators. The chromatic length parameter (eq. 2) has been defined to give an analytical description of the beam emittance degradation in free space propagation. When the $L_c$ is many times longer than a TL that have to be designed, the emittance is constant and classical matching approach are successful, differently when the $L_c$ is shorter or comparable with the TL path length the problem is much more complex and classical methods cannot properly work. The effect of one optic (e.g. a quad) can rise the $L_c$ accelerating the emittance degradation that in turn correlates with following optics. These results in a non-linear problem that are typically well solved by GA as demonstrated in this study.

GIOTTO has been improved in order to deal with WRS problems, as the design of a transfer line able to match to the undulator a plasma accelerated beam. The key of this improvement is the adoption of Lorentzian function as part of the fitness function.

We showed 4 alternative lines able to match to undulator the given beam preserving the normalized transverse emittance. In particular, one can deal with different beam energies with the adoption of tunable PM Quads.

In the most promising line solution have been checked to work with higher particle statistics and space charge effects.